# A simple model for an internal wave spectrum dominated by non-linear interactions

## by Hans van Haren[1*], Leo Maas[2]

[1]**Royal Netherlands Institute for Sea Research (NIOZ), P.O. Box 59, 1790 AB Den Burg, the Netherlands**
*Corresponding author. e-mail: **hans.van.haren@nioz.nl**
[2]Institute for Marine and Atmospheric Research (IMAU), University of Utrecht, P.O. Box 80011, 3508 TA Utrecht, the Netherlands

ABSTRACT

Ocean motions at frequencies of the internal wave band are generally associated with freely propagating waves that are supported by stable vertical stratification in density. Previous analyses of yearlong current observations from the Bay of Biscay showed that a finestructure of semidiurnal tidal and near-inertial higher harmonics fills the spectrum. Here, a simple model is presented of forced nondispersive motions with forward energy cascade. The model fits the spectral shape of higher harmonics well within statistical significance and shows that such interactions imply maximum wave steepness in a balance between forcing and turbulent mixing. The single fitting parameter is related to the barotropic tidal flow speed, which thereby sets a nonlinear limit to baroclinic current scales without generating non-linear higher harmonics directly.

*Keywords: internal wave observations; Bay of Biscay; non-linear higher harmonics; advection model; forward cascade*

# 1. Introduction

Following observations (Pinkel et al., 1987; Mihaly et al., 1998; van Haren et al., 1999) and numerical modeling (e.g., Xing and Davies, 2002; Pichon et al., 2013) the oceanic internal wave band (IWB) is not always a smooth broadband spectrum, but it can be dominated by peaks associated with near-inertial and/or semidiurnal tidal motions and their higher harmonics.

In theory, freely propagating internal gravity waves can exist in the IWB frequency ($\sigma$) band between $|f(\varphi)| < \sigma < N(z)$, showing the depth (-z) dependent buoyancy frequency $N(z) = (-g(d\ln\rho/dz+g/c_s^2))^{1/2}$, $N \gg f$, where g is acceleration of gravity, $c_s$ the speed of sound describing compressibility effects, and latitudinal ($\varphi$) dependence is indicated of inertial frequency $f(\varphi) = 2\Omega\sin\varphi$, twice the local vertical component of the Earth's rotation vector $\mathbf{\Omega}$ (e.g., LeBlond and Mysak, 1978). Suggestions have been made (Mihaly et al., 1999; Xing and Davies, 2002) that near-inertial motions are important for transfer of energy inside the IWB through non-linear interaction with semidiurnal tidal motions. Higher tidal and inertial-tidal harmonics are not only observed in shallow seas (van Haren et al., 1999), but also in the deep ocean near and away from topography (Mihaly et al., 1998; van Haren et al., 2002). These observations seemed to confirm the hypothesis that (breaking) non-linear internal waves may be important for diapycnal mixing (Müller and Briscoe, 1999).

In this paper a simple model is proposed describing non-linear interactions that generate such internal wave band higher harmonics peaks and fit the spectral shape of yearlong current meter observations from the deep Bay of Biscay. The model follows theoretical suggestions (Phillips, 1977) and adds to continuous smooth spectra (Garrett and Munk, 1972) describing a symmetric and isotropic linear wave field. Previously, no model existed for the spectral shape of non-linear higher harmonics, although there were studies about relative importance of such constituents in tidal context, mainly for shallow seas (Dronkers, 1964; Parker, 1991; Pingree and Maddock, 1978). Hence, their naming as 'shallow water--tidal--constituents'. Here, we consider non-linear advection, which occurs in the equations for momentum and buoyancy, to describe deep-ocean internal wave higher harmonic interaction frequencies.

**2. Observations**

Currents were evaluated from two moorings deployed in the Bay of Biscay NE-Atlantic Ocean during 11 months, above the continental slope at 46°39′ N, 05°29′ W (water depth H= 2450 m) and above the abyssal plain at 45°48′ N, 06°50′ W (H = 4810 m), see Fig. 1. The focus is on data from the uppermost Aanderaa RCM-8 single-point current meter at 1000 m above the bottom in each mooring. This distance above the seafloor is well above any internal wave breaking at the local seafloor slope. Horizontally, it is at least 10 km from topography, and the deepest site is more than 100 km from the foot of the continental slope. The rough continental slope may generate internal (tidal) waves and possibly focus them by reflection. It is unknown however, how far from the slope its effects are sensed. Numerical modelling suggests the spread of internal tides throughout the Bay of Biscay (e.g., Pichon et al., 2013).

From a few CTD density profiles, obtained near the moorings, stratification was estimated $N(z) = (1\pm0.5)(20 + 0.0034z)$ cpd, $-4480 < z < -2740$ m (frequency was calculated in cycles per day, 1 cpd = $2\pi/86400$ s$^{-1}$). This depth dependence changed abruptly above 2740 m. At 1500 m, $N \approx 28$ cpd.

Tidal harmonic analysis (Dronkers, 1964) was used to split a highly deterministic narrowband large-scale signal, here termed 'barotropic' signal, from the remainder ('baroclinic' or internal, difference signal). Because we use current meters, the barotropic signal represents a time-coherent signal at a limited number of semidiurnal constituents. At semidiurnal lunar tidal $M_2$ the barotropic signal is relatively largest and about twice the value of the baroclinic signal.

Observed kinetic energy spectra $P_{KE}(\sigma)$ revealed larger energy at shallower depth (where N is larger), except at f (Fig. 2). For the entire IWB-band relatively largest energy was found at localized frequencies associated with inertial and semidiurnal tidal motions (indicated as f and (lunar) $M_2$, respectively) and higher harmonics (indicated as $M_4$, $M_6$,…and $M_2+f$, $M_4+f$…). These energy peaks exceeded the spectral continuum that sloped with frequency like $P_{KE} \sim \sigma^{-1}$, for $f < \sigma < 7$ to 10 cpd. At higher frequencies, the continuum sloped steeper. Higher harmonics were observed above their respective continuum levels up to $M_{10}$ at the deeper site and up to $M_{16}$ at the shallower site. Motions that not exceeded, or were included in, the continuum, were found at, e.g., $2f$, $3f$, $M_2+2f$ (van Haren et al., 2002).

When the kinetic energy was large at f (deepest mooring, small N), energy at f-interaction frequencies (e.g. $M_2+f$) showed a spectral fall-off rate with frequency like $P_{KE} \sim \sigma^{-3}$, which was a typical fall-off rate, or even steeper, for higher tidal harmonics, see red plusses in Fig. 2. When energy at f was reduced (shallowest mooring, large N), energy at $M_n+f$, n=2, 4,…, scaled like $\sim\sigma^{-2}$. When smoothed strongly, the latter records showed overall spectral fall-off rate close to $\sim\sigma^{-2}$ for $f < \sigma < N$ (van Haren et al., 2002). For the deepest mooring with relatively large inertial-tidal and tidal harmonics, the heavily smoothed overall spectral fall-off rate was faster $\sim\sigma^{-3}$. This is significantly steeper than the canonical fall-off rate $P_{KE} \sim \sigma^p$, -2.5 < p < -1.5 for open-ocean internal waves (Garrett and Munk, 1972).

**3. Spectral model for internal wave higher harmonics**

Formally, free plane internal waves in a continuously stratified medium are governed by linear equations and the non-linear advection terms exactly cancel in the equations for vorticity and buoyancy. This follows from the absence of variations in the planes of equal phase, and which distinguishes internal from surface waves. However, in reality non-linearity is not expected to vanish. Like surface waves internal waves may manifest themselves as displacement waves on layers of enhanced stratification. Such displacement waves may grow up to highly non-linear shock waves (Platzman, 1964). Due to variations in source, for example at different sites and due to slowly varying stratification, internal waves propagate in groups of limited size which provides their intermittent character, and non-linearity may be imperative to prevent dispersion (Thorpe, 1999).

Sofar, no general solution has been given to describe non-linear internal wave motions using the full set of governing equations (Shrira, 1981). The mathematics is too complex. However, considering internal waves occurring intermittently and in groups, a simple model is suggested that describes the energy spectrum of higher harmonics, somewhat along the lines of non-linear wave deformation like in shock waves (Platzman, 1964).

Consider the discrete kinetic energy spectrum $P_j(\sigma) = \frac{1}{2}\hat{U}_j\hat{U}_j^*$, the asterisk denoting complex conjugate, of current components $\hat{u}_j = \hat{U}_j\exp[i(k_j\xi-\sigma_j t)]$ with frequency $\sigma_j$, wavenumber $k_j$ and amplitude $\hat{U}_j$, j being a positive integer starting from the fundamental baroclinic component, $i^2 = -1$. The coordinate $\xi$ is in the oblique energy propagation direction of the deformed wave, parallel to the phase lines, and it is assumed to indicate the direction of the largest current component $\hat{u}$. The latter may have a relatively large component in the vertical direction.

In the model discrete spectrum only those parts are considered that are entirely governed by advection in only one direction ($\xi$) and no advection perpendicular to this direction, expressing forced non-linear interactions between linear motions that result in bound non-freely-propagating motions. As motions at interaction frequencies $\sigma \gg f$ are considered, rotation is neglected. A pressure gradient forcing is assumed being linear and only governing the fundamental tidal components. Diffusion is neglected. Excluding advection by a 'barotropic' surface tidal current $\hat{u}_0$ ($\hat{U}_0$, $\sigma_0 = \sigma_1$, $k_0$) as we consider motions within a barotropic oscillating system and because the wavenumber $k_0 \ll k_1$, we examined the following 'cartoon' model, apparently capturing our observations,

$$\frac{\partial \hat{u}_j}{\partial t} = -\sum_{i=1}^{j-1} \hat{u}_i \frac{\partial \hat{u}_{j-i}}{\partial \xi}. \qquad (1)$$

In general, the resonance conditions for frequency and wavenumber imply that the compound wave does not satisfy the dispersion relation. Here, only non-resonant higher harmonics $\sigma_j = j\sigma$ and $k_j = jk$ are considered. This assumes a forward cascade of energy from a source at a fundamental 'baroclinic' internal (tidal, inertial) constituent ($\sigma_1$, $k_1$, $\hat{U}_1$) ≡ ($\sigma$, k, $\hat{U}$), so that (1) simplifies to the equation,

$$\hat{U}_j = \left(\frac{k}{\sigma}\right)\sum_{i=1}^{j-1}\frac{j-i}{j}\hat{U}_i\hat{U}_{j-i} \equiv q_j\gamma^{j-1}\hat{U}, \qquad (2)$$

recursively defining factors $q_j = \frac{1}{2}, \frac{1}{2}, \frac{5}{8}, \frac{7}{8}, \frac{21}{16}, ...$ for j = 2, 3, 4, 5, ..., etcetera, and depending only on parameter $\gamma = \hat{U}/c$, with c=$\sigma$/k denoting the phase-speed of the fundamental harmonic. A constant c ($c_j \equiv \sigma_j/k_j = c$) implies non-dispersive wave growth. The model (2) contrasts with the shock wave model by Platzman (1964), which also includes a backward cascade.

From (2) a consistent model spectrum $P_m(\sigma) = \Sigma_j P_j(\sigma)$ is obtained fitting the observed kinetic energy at discrete frequencies $P_{df}(\sigma)$ for $\gamma$ = 0.48±0.05 under the conditions |log($P_{df}(\sigma)/P_m(\sigma)$)| < 10% and $\sigma$ < N (Fig. 2). The small standard deviation, which is well within 95% statistical significance of the spectra, expresses the sensitivity of the model to $\gamma$, with the surprising result that $\hat{U}_0$ = (0.96±0.11)c, using observed $\hat{U}_0/\hat{U}$ = 2.0±0.15 following harmonic analysis splitting the original signal into semidiurnal time-coherent signal and its remainder baroclinic signal. As a result, large scale barotropic $\hat{U}_0$ is found setting a non-linear limit that determines baroclinic $\hat{U}$-length scales, whilst not generating non-linear constituents directly. This can be seen as for $k_0 \to 0$ barotropic advection yields only forced, non-resonant, dispersive ($\sigma_j$ = j$\sigma$, $k_j$ = k) harmonics $\hat{U}_j = (\hat{U}_0/c)^{j-1}\hat{U}/j!$. These constituents show a much faster energy drop with frequency than $\sigma^{-3}$ (for $\sigma$ > $M_4$) using the same $\gamma$.

The model (2) is subsequently applied for distinct quasi-deterministic f and $M_2$ frequencies. However, it equally applies for all neighbouring frequencies. In fact, the shape of the primary inertial-semidiurnal f/$M_2$ band is seen to be transposed to the $M_2$+f/$M_4$ band following (2), and similarly to higher frequency bands. Hereby, the shape-shrink in frequency is attributable to log-log plotting and the variance-shrink to (2) that equally affects all frequencies in the primary band. The splitting of energy to neighbouring frequencies is in part due to interactions with the slowly varying stratification background (van Haren, 2016).

**4. Discussion**

The relatively largest tidal and inertial-tidal higher harmonics are observed at the deep site, more than 100 km from the continental slope. It may be questioned whether these higher harmonics are reminiscent of open-ocean non-linear interactions or of the effects of sloping underwater topography reaching that far into the deep ocean. Observations are lacking of detailing turbulence, but numerical internal tie modelling demonstrates multiple interactions at such distance from topography (Pichon et al., 2013). The non-linearity as in observed higher harmonics suggests that turbulence may not be negligible.

This is because similarity of particle displacement speed and phase speed of semidiurnal signal suggests a gradient Richardson number $Ri \approx 1$, or a transition from weak wave-wave interaction to, presumably strong, stratified turbulence (Phillips, 1977; D'Asaro and Lien, 2000). In shallow seas, this leads to marginal stability and associated turbulent diapycnal exchange that is sufficient for nutrient replenishment into the photic zone (van Haren et al., 1999). Here, it is interpreted as saturation of non-linear gradients balanced by mixing parameterized by $\gamma$. As $\gamma=\pi L/\lambda$ ($L=2\hat{U}/\sigma$ denoting particle excursion length), the model results imply horizontal wavelength of about 5 km of the fundamental constituents and particle speeds of fundamental constituents of 0.05 m s$^{-1}$ as observed, for both inertial-tidal and tidal higher harmonics. Because $\gamma$ was found independent of $N$, this length-scale may be fundamental for baroclinic non-linear transfer via advection. Perhaps, this simple model invites future clarification of a balance between internal wave forcing and diapycnal mixing in more sophisticated non-linear internal wave models.

**Acknowledgements**

The assistance of the crew of the R/V Pelagia was greatly enjoyed. The Bay of Biscay Boundary (TripeB; Hendrik van Aken) project was supported by grants from the Netherlands organization for the advancement of scientific research, NWO.

**Disclosure statement**

No potential conflict of interest was reported by the authors.

**References**

D'Asaro, M. and Lien, R. 2000. The wave-turbulence transition for stratified flows. *J. Phys. Oceanogr.* **30**, 1669-1678.

Dronkers, J. J. 1964. *Tidal Computations in Rivers and Coastal Waters,* Amsterdam, NL: North Holland Publishing Company, 518 pp.

Garrett, C. J. R. and Munk W. H. 1972. Space-time scales of internal waves. *Geophys. Fluid Dyn.* **3**, 225-264, 1972.

LeBlond, P. H. and Mysak, L. A. 1978. *Waves in the Ocean*. New York, USA: Elsevier, 602 pp.

Mihaly, S. F., Thomson, R. E. and Rabinovich, A. B. 1998. Evidence for non-linear interaction between internal waves of inertial and semidiurnal frequency. *Geophys. Res. Lett.* **25**, 1205-1208.

Müller, P. and Briscoe, M. 1999. Diapycnal mixing and internal waves. In: *Dynamics of oceanic internal gravity waves, II. Proceedings 'Aha Huliko'a Hawaiian Winter Workshop* (ed. P. Müller and D. Henderson), Hawaii, USA: SOEST, pp. 289-294.

Parker, B. B. (ed.) 1991. *Tidal hydrodynamics.* New York, USA: John Wiley & Sons, 883 pp.

Phillips, O. M. 1977. *The Dynamics of the upper Ocean* (2$^{nd}$ Ed.). Cambridge, UK: Cambridge University Press, 336 pp.

Pichon, C., Morel, Y., Baraille, R. and L. S. Quaresma, 2013. Internal tide interactions in the Bay of Biscay: Observations and modelling. *J. Mar. Sys*. **109-110**, S26-S44.

Pingree, R. D. and Maddock, L. 1978. The $M_4$ tide in the English Channel derived from a non-linear numerical model of the $M_2$ tide. *Deep-Sea Res.* 26, 53-68.

Pinkel, R., Plueddemann, A. and Williams, R. 1987. Internal wave observations from FLIP in MILDEX. *J. Phys. Oceanogr*. **17**, 1737-1757.

Platzman, G. W. 1964. An exact integral of complete spectral equations for unsteady one-dimensional flow. *Tellus* **16**, 422-431.

Shrira, V. I. 1981. On the propagation of a three-dimensional packet of weakly non-linear internal gravity waves. *Int. J. Non-lin. Mech.* **16**, 129-138.

Thorpe, S. A. 1999. On internal wave groups. *J. Phys. Oceanogr.* **29**, 1085-1095.

van Haren, H. 2016. Do deep-ocean kinetic energy spectra represent deterministic or stochastic signals? *J. Geophys. Res*. **121**, 240-251, doi:10.1002/2015JC011204.

van Haren, H., Maas, L., Zimmerman, J. T. F., Ridderinkhof, H. and Malschaert, H. 1999. Strong inertial currents and marginal internal wave stability in the central North Sea. *Geophys. Res. Lett*. **26**, 2993-2996.

van Haren, H., Maas, L. and van Aken, H. 2002. Construction of 'internal wave' spectrum near a continental slope. *Geophys. Res. Lett*. **29(12)**, 10.1029/2001GL014341.

van Haren, H., Maas, L. R. M. and Gerkema, T. 2010. Patchiness in internal tidal beams. *J. Mar. Res*. **68**, 237-257.

Xing, J. and Davies, A. M. 2002. Processes influencing the non-linear interaction between inertial oscillations, near inertial internal waves and internal tides. *Geophys. Res. Lett*. **29(5)**, 10.1029/2001GL014199.

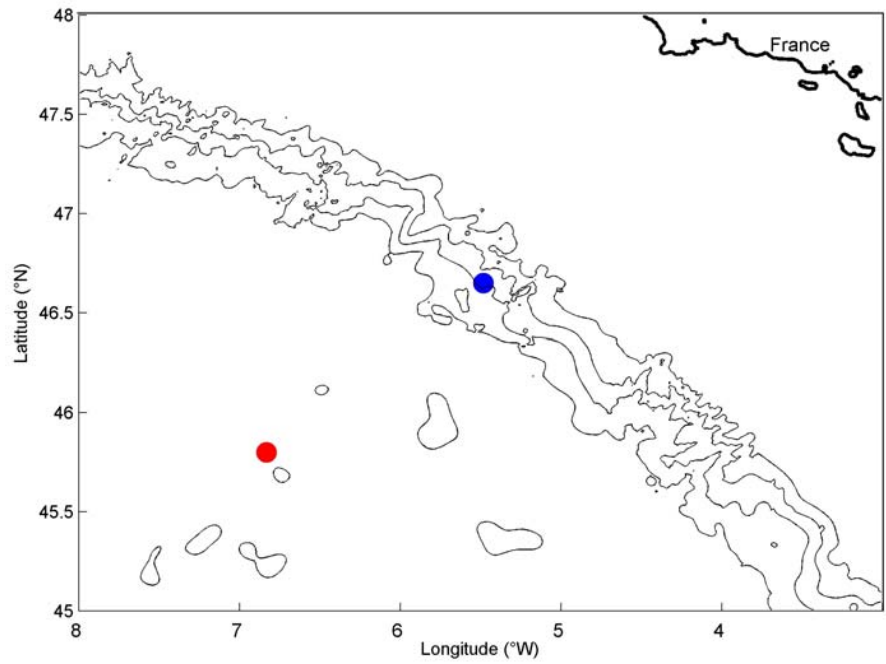

*Fig. 1.* Current meter mooring sites in the Bay of Biscay with black contours of topography every 1000 m water depth. The coloured dots correspond with the spectra in Fig. 2.

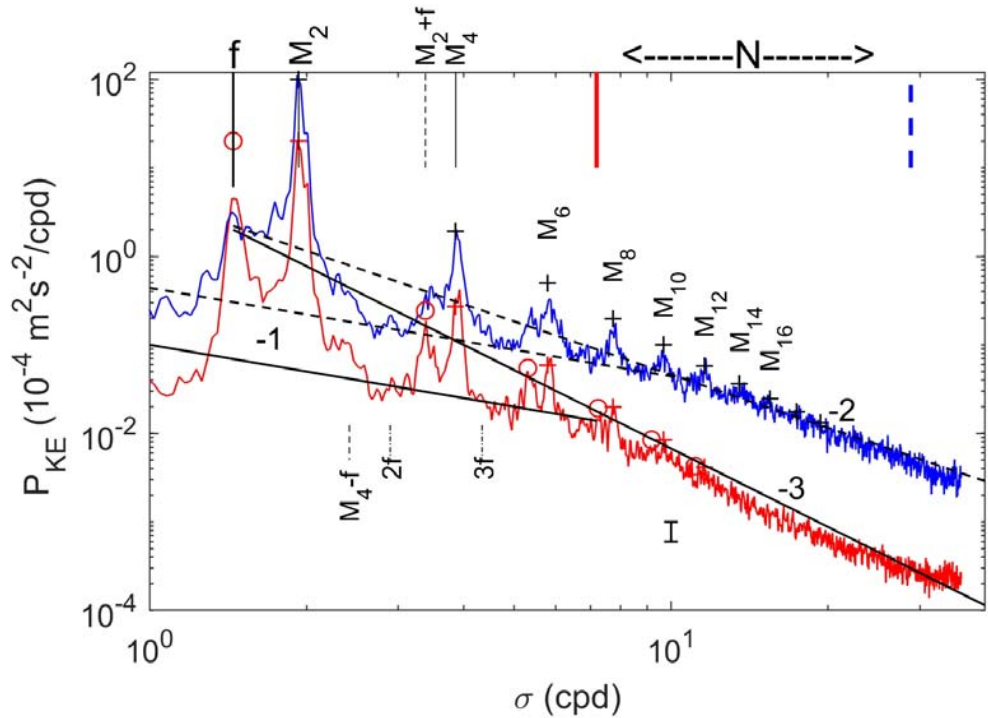

*Fig. 2*. Kinetic energy spectra from 11 months of current meter observations at 1000 m above the seafloor in H=4810 m water depth (red) and H=2450 m (blue). Spectra were moderately smoothed ($\nu \approx 30$ df) and not offset vertically. The difference in energy levels between the spectra corresponded to the difference in N(z), which variation is indicated between the vertical bars in the top-right corner. This corresponds with the vertical distance between the sloping lines at fall-off rates $\sigma^{-1}$ (solid and dashed corresponding to red and blue spectra, respectively). Constant slopes in log-log plot are indicated "-1,-2,-3" representing $\sigma^{-1}$, $\sigma^{-2}$, $\sigma^{-3}$, respectively. Spectra of model (2) are superposed for observed barotropic and baroclinic fundamental tidal amplitudes and fitting parameter γ. Three model examples are given, one for inertial-tidal-interactions (o) and two for tidal-interactions (+). They fit well the observed energy levels for nearly the same γ (see text). In all cases, reference amplitude is the barotropic $M_2$ current amplitude, indicated at f and $M_2$ (leftmost o, +). Baroclinic $M_2$-variance are a quarter of peak $M_2$-values.